\documentstyle[prl,aps]{revtex} \def\narrowtext{} \tighten\twocolumn

\begin{document}
\draft
\title{
\begin{minipage}[t]{7.0in}
\scriptsize
\begin{quote}
\leftline{{\it J. Appl. Phys.} {\bf 91}, 7508 (2002).}
\raggedleft{\rm OUTP-01-02S}\\
\raggedleft {\rm cond-mat/0110322}
\end{quote}
\end{minipage}
\medskip
\\Spin Waves, Phase Separation, and Interphase Boundaries
in Double Exchange Magnets.}
\author{D. I. Golosov\thanks{E-mail: golosov@thphys.ox.ac.uk}}
\address{Theoretical Physics, Oxford University, 1 Keble Rd. Oxford
OX1 3NP United Kingdom}  
\address{%
\begin{minipage}[t]{6.0in}
\begin{abstract}
We study a classical double exchange magnet with  direct antiferromagnetic 
superexchange coupling, J, between the localized spins. It is shown that 
the de-stabilization of the ferromagnetic ground state with increasing J
leads to phase separation; the latter always preempts the spin-wave
instability (softening of the magnon spectrum). It is also found that
the boundaries separating the ferromagnetic and antiferromagnetic areas of 
the sample tend to be abrupt. 
\typeout{polish abstract}
\end{abstract}
\end{minipage}}

\maketitle
\narrowtext
\newpage

The phenomenon of phase separation in double exchange (or $s$-$d$
exchange) magnets has been discussed in the literature for a long time
\cite{Nagaev72}. Presently it attracts attention of both theorists and
experimentalists investigating the properties of doped manganese 
oxides which exhibit colossal magnetoresistance (CMR) \cite{reviewDagotto}.

One of the simplest yet physically relevant scenarios of phase
separation takes place in a  
single-orbital double exchange magnet 
in the presence of a sufficiently strong antiferromagnetic 
superexchange coupling $J$ between the neighbouring localized spins.
Some of the earlier articles\cite{Nagaev96,prb98,Kagan} noted
that the thermodynamic instability of de Gennes' canted phase
(originally\cite{deGennes} thought to emerge as an outcome of double
exchange -- superexchange competition) at small values of band filling 
$x$ indicates that
this phase is unstable with respect to phase
separation. Others\cite{Arovas,Yamanaka} used mean field and
variational approaches to minimise the ground state energy and obtained
rich phase diagrams. The
numerical studies\cite{reviewDagotto,Aliaga} 
suggest that phase separation occurs at all values of $x$, provided that
the value of $J$ is sufficiently large. The systematic physical
understanding of phase separation in these
systems is however still incomplete. In the present article, we will 
address some aspects of this deficiency, paying special attention
to the relationship between phase separation and spin waves, as well
as to the features of the 
interphase boundaries, which are likely to affect both equilibrium and
transport properties of the system.   

We consider a model defined by the Hamiltonian
\begin{eqnarray}
{\cal H}=&&-\frac{t}{2} \sum_{\langle i,j \rangle,\alpha} 
\left(c^\dagger_{i \alpha}c_{j \alpha} +c^\dagger_{j \alpha}c_{i
\alpha}\right) - 
 \frac{J_H}{2S} \sum_{i, \alpha, \beta} \vec{S}_i
\vec{\sigma}^{\alpha \beta} c^\dagger_{i\alpha} c_{i\beta}+ \nonumber \\
&& +\frac{J}{S^2}\sum_{\langle i, j \rangle} \vec{S}_i \vec{S}_j\,.
\label{eq:Ham}
\end{eqnarray}
Here  $c_{j \alpha}$  (with $\alpha= \uparrow,\downarrow$) 
are the electron annihilation operators,
$\vec{S}_i$ are the operators of the core (localized) spins
located at the sites of a square lattice, and
the vector $\vec{\sigma}^{\alpha \beta}$ is composed of Pauli
matrices. We assume that ionic spins are classical ($S
\rightarrow \infty$), and  the Hund's
rule exchange coupling is large ($J_H \rightarrow \infty$).
Under these assumptions, there is only one
carrier spin state available at each site, and the hopping coefficient
is effectively renormalized\cite{deGennes,Anderson} by
a factor $(S^2+\vec{S}_i \cdot \vec{S}_j)^{1/2}/(S\sqrt{2})$. 
Throughout the paper we use units in which hopping $t$
and the lattice spacing are equal to unity, and we consider the
$T=0$ case. While technically our
treatment is restricted to the
two-dimensional model, we expect that the conclusions 
remain qualitatively valid in three dimensions.

When the value of superexchange $J$ is small, the ground state of the
system is ferromagnetic (FM), and the carrier spectrum is given by\cite{omit} 
$ \epsilon_{\vec{k}}=- \cos k_x - \cos k_y$. The total energy of the
conduction band, $E=\int_{-2}^{\mu} \epsilon \nu(\epsilon)
d\epsilon$ (where $\nu(\epsilon)$ is the corresponding density of
states) is determined by the value of chemical potential $\mu$.

As the value of $J$ increases, the ferromagnetic ground state of
a double exchange magnet is eventually destabilized, which  can be
seen in particular from the fact that at a certain $J_{sw}=|E|/8$ the spin wave
spectrum of  FM phase
$\omega_{\vec{p}}=(-2J+|E|/4)(2+\epsilon_{\vec{p}})/S$
(see Ref. \cite{jap00} and references therein), 
softens and then turns negative. The nature of
emerging phase has not been clarified by a spin wave theory
investigation\cite{jap00}, although it was demonstrated that the
homogeneous de Gennes' canted state can never be stabilized.
We will now show that in reality this spin wave instability of a
homogeneous, fully saturated  FM state can never be reached as the
system becomes
unstable with respect to phase separation already at $J<J_{sw}$. 
This conclusion will be verified for all values of band filling
$x$, and the emerging new phases will be discussed. 

As the value of $J$ increases, the homogeneous FM state with a
chemical potential $\mu=\mu_{FM}(x)$ is destabilized with respect to phase
separation whenever its thermodynamic potential, $\Omega_{FM}=E-\mu x
+2 J$,
becomes larger than the thermodynamic potential of some other phase,
calculated at the same value of $\mu$. Further increase of $J$ will
result in an increase of the volume occupied by this new phase at the
expense of the FM area; the combined change of values of $J$ and
$\mu$ will eventually lead to another instability or phase
transition, etc. Thus in order to determine  the stability region for
the homogeneous FM state at a given value of $x$, one has to identify 
the phase P, for which the condition
$\Omega_P(\mu_{FM}(x),J_P)=\Omega_{FM}(\mu_{FM} (x),J_P)$ is reached 
at the lowest 
possible $J_P$. In doing so, we are restricted by the usual limitations
of variational approach\cite{zoo}. Results are summarized in Fig.
\ref{fig:phasesepar}; note that for any\cite{simm} value of $x$, there is at
least one phase corresponding to a value of $J_P$ smaller than
$J_{sw}(x)$. This allows us to 
conclude that {\it phase separation}, rather than a second-order phase
transition associated with the spin-wave instability at $J=J_{sw}$,
{\it is
the generic outcome of double exchange -- superexchange competition}. 

We consider the following phases:

\noindent {\bf G-type AFM} (G-AFM) phase with checkerboard
spin ordering. The thermodynamic potential is given by\cite{negative}
$\Omega_G=-2 J$.
Given our selection of phases, the G-AFM phase corresponds to the
lowest value of $J_{P}$ at $0<x<0.245$.  

\noindent {\bf A-type AFM} (A-AFM) phase characterized by the wave
vector $\{\pi,0\}$. We find that this phase with thermodynamic
potential 
$\Omega_A=-\{\sqrt{1-\mu^2}+\mu\, {\rm arc} \sin \mu +\pi
\mu/2 \} \theta(1-\mu^2)/\pi$
yields the lowest value\cite{flux} of $J_{P}$ at $0.37<x<0.5$. 

\noindent {\bf Island phases} represent a generalization of
one-dimensional island phases \cite{island}, and were already
found in  the numerical investigations of both two- and three-dimensional
cases\cite{Aliaga}. The thermodynamic potential for the 2x2 island
phase (see Fig. \ref{fig:phasesepar} {\it b}) is given by
$\Omega_2=-(\mu+1)\theta(1-\mu^2)/4$, 
and this phase corresponds to the lowest $J_{P}(x)$ for $0.293<x<0.335$.
Similarly, at $0.335<x<0.37$ the lowest value of $J_P$ corresponds to the 3x3
island phase.

\noindent {\bf Chain phase} (see Fig. \ref{fig:phasesepar} {\it b}) is
proposed here for the first time. Like in  A-AFM phase, 
carriers in the chain phase are not completely localized: they can
travel along the braided paths parallel to the direction of the chains.
We find 
\[\Omega_{ch}=-\frac{2}{3 \pi}\left\{ \sqrt{2-\mu^2}+\mu\, {\rm arc} \sin
\frac{\mu}{\sqrt{2}}+\frac{\pi}{2} \mu \right\} \theta (2-\mu^2)+\frac{2 J}{3}\,,\]
and the corresponding value of $J_{P}$ is the lowest for $0.245<x<0.293$.

Phase separation obviously leads to formation of interphase
boundaries separating different areas of the sample. It is very
unlikely that these are smooth, Bloch-like walls with gradually increasing
magnetization and decreasing staggered magnetization -- indeed, this
would inevitably involve the appearance of (unstable\cite{jap00})
canted spin ordering in the intermediate region. The extreme opposite case of
an abrupt interphase boundary corresponds to a complete change of magnetic
order over one lattice link (see Fig. \ref{fig:wall} {\it a}); for the case of
phase separation into the A-AFM and FM phases, such
boundaries should run along a crystal axis, and for the G-AFM or chain
phases -- in a diagonal direction. Short of a rigorous proof that
these abrupt walls represent the optimal configuration, we will probe
their stability with respect to small ``smearing'' perturbations \cite{smear} 
shown
schematically in Fig. \ref{fig:wall} {\it a}. The energy cost (per
unit length) of an abrupt wall separating the FM phase form another
phase P can be written in the form
\begin{equation}
\delta \Omega =W_P(\mu,J)+{\cal A}_P(\mu,J) \varphi^2 + {\cal B}_P
\psi^2\,\,,\,\,\,\,\,\,\varphi,\psi \ll 1\,.
\label{eq:boundary}
\end{equation}
The evaluation of $W$, ${\cal A}$, and ${\cal B}$ is sketched
in the Appendix.

The values of the coefficients ${\cal A}_P(\mu_FM(x),J_P(x))$ and 
${\cal B}_P(\mu_{FM}(x),J_P(x))$ are plotted in Fig. \ref{fig:wall} {\it b}.
We see that the energy cost of perturbing an abrupt FM -- G-AFM 
boundary is always positive throughout the region where
the value $J_{G}(x)$ is the lowest among the $J_P$'s
evaluated above (see Fig. \ref{fig:phasesepar} {\it a}). A similar
situation occurs for an FM -- A-AFM boundary. This represents
rather convincing evidence to the effect that {\it the boundaries
separating the FM and AFM areas are abrupt}. This conclusion  is not
altogether unexpected as it is known, {\it e.g.},
that small polarons
propagating in a doped antiferromagnet have no tails\cite{Auerbach},
{\it i.e.}, are ``abrupt''.

We speculate
that the instability of an abrupt chain -- FM wall at $x>0.265$, seen
from Fig. \ref{fig:wall} {\it b}, may
indicate the existence of another, so far unidentified phase $P$ with a
lower critical value of superexchange, $J_P<J_{chain}$.

The author takes  pleasure in thanking J. T. Chalker, A. Auerbach, and
A. M. Tsvelik  for enlightening and motivating discussions. This
work was supported by EPSRC under grant GR/J78327.

\appendix
\section{}

Here we outline the main steps in the calculation of 
coefficients on the r.\ h.\ s.\ of Eq.(\ref{eq:boundary}). Since it is
assumed that the interphase boundary is  perfectly straight, the
variables separate, and one is left with an effective one-dimensional
problem (Fig. {\ref{fig:wall} {\it a}). For simplicity,  we will
describe the case of a boundary between the FM and G-AFM or A-AFM
phase; the chain phase is treated similarly.
Carrier 
hopping in the direction perpendicular to the boundary vanishes in
the AFM phases, so the presence of a bulk AFM phase to the left of the
boundary does not affect the electron wave functions in the FM part of
the sample.
The electronic contribution to
the energy cost of a boundary
is due solely to the associated change in the wave functions in the FM 
part of the sample (these do not extend across the boundary). 
Thus, the energy of the interphase boundary  is closely
related to the energy of an abrupt FM domain wall depicted in 
Fig. {\ref{fig:wall} {\it c} (top); indeed, it is obvious that the
latter energy is equal to 
$2W_P+{\cal A}_P \varphi^2 + {\cal B}_P \psi^2$. The origin of the
prefactor $2$ in the first term is clarified in  Fig. \ref{fig:wall}
{\it c} (bottom), showing {\it two} FM-AFM boundaries (only one of which 
is perturbed at $\varphi, \psi \neq 0$), separated by three simple chains of
AFM between them. Generally, the energy difference between such ``stripe''
configurations and the
domain wall shown in  Fig. {\ref{fig:wall} {\it c} (top) 
is equal to $\Omega_{AFM}-\Omega_{FM}$ times the number
of inserted  chains of AFM\cite{factor}, {\it i.e.}
is entirely due to the insertion of the bulk AFM phase. 
In turn, the abrupt domain wall
is a {\it local} perturbation of the FM order, and its energy can be
conveniently evaluated with the help of the Lifshits--Krein trace formalism 
(see Ref. \cite{prb98}, and references therein). We find\cite{misprint}:
\begin{eqnarray}
W_G &=&\frac{2^{-3/2}}{\pi}\left\{\sqrt{4-\mu^2}+\mu\, {\rm
arc} \cos \frac{|\mu|}{2}\right\}-2\sqrt{2}J-\nonumber \\
&&-\frac{\Omega_{G}-\Omega_{FM}}{2\sqrt{2}}\,,
\nonumber\\ 
W_A &=&\frac{\sqrt{-2\mu-\mu^2}}{4 \pi}+ \frac{1+\mu}{4\pi }{\rm
arc} \cos(|\mu|-1)-\nonumber\\
&&-\theta(1+\mu) \cdot\frac{1}{2\pi}\left[\mu{\rm
arc} \cos(-\mu)+\sqrt{1-\mu^2}\right] - \nonumber \\
&&-2J-\frac{1}{2}(\Omega_{A}-\Omega_{FM}) \,,
\nonumber
\end{eqnarray}
and rather lengthy expressions for ${\cal A}_P$ and ${\cal B}_P$.
At $x \ll 1$, we obtain $W_G \approx
-\sqrt{2}J+(\mu+2)^{3/2}/(3\sqrt{2}\pi)$;  the 3D counterpart of this
expression was used in Ref. \cite{Nagaev72}.

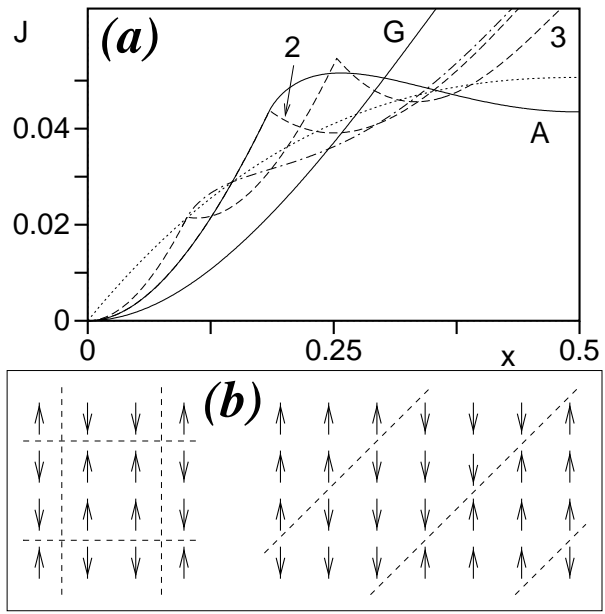
\begin{figure}
\caption{{\it(a)} Values of superexchange $J$ corresponding to 
the instabilities of  FM order. The solid line labeled ``G'' (``A'')
represents the phase separation into G-type (A-type) AFM and FM
phases, whereas the dashed line labeled ``2'' (``3'') corresponds to
phase separation into 2x2 (3x3) island phase and FM phase. Phase
separation into FM and chain phases (dashed-dotted line) may also be
possible at certain values of band filling $x$. Note, however, that
the spin-wave instability at $J=J_{sw}(x)$ (dotted line) is always
circumvented by phase separation.   {\it(b)} Spin ordering in 
the 2x2 island (left) and chain (right) phases.} 
\label{fig:phasesepar}
\end{figure}

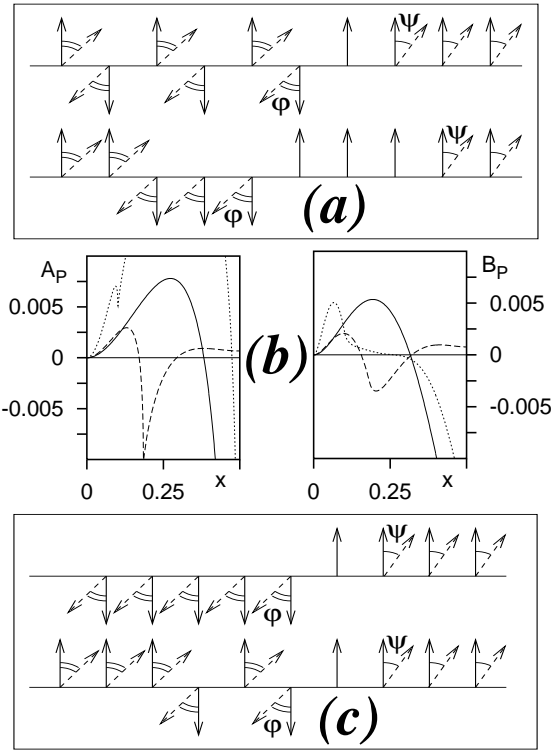
\begin{figure}
\caption{{\it (a)}. Spin arrangements near the abrupt boundaries between 
FM and G-AFM or A-AFM phases (top) and between FM and chain phases
(bottom). Each arrow represents a FM chain parallel to the boundary;
interchain distance is equal to unity for the FM -- A-AFM boundary,
and to $1/\sqrt{2}$ for the FM -- G-AFM and FM -- chain cases. Dashed
arrows correspond to perturbed boundaries with $\varphi, \psi \neq 0$.
{\it (b)} The values of coefficients ${\cal A}_P(\mu_{FM}(x),J_P(x))$ (left)
and ${\cal B}_P(\mu_{FM}(x),J_P(x))$  (right) 
for the boundaries between FM and G-AFM (solid), A-AFM (dashed), or
chain (dotted) phases. See Eq. (\ref{eq:boundary}). {\it (c)}
An abrupt domain wall in the FM state -- see Appendix.}
\label{fig:wall}
\end{figure}

\begin{references}
\bibitem{omit} Here
and below, we omit a large constant term proportional to $J_H$.
\bibitem{zoo} In addition, we expect that at intermediate doping
levels, the energy differences
between various phases are very sensitive to details of both bandstructure
and exchange interaction.
\bibitem{simm} The properties of this model at $J_H,S \rightarrow \infty$
are symmetric with
respect to $x=0.5$.
\bibitem{negative} Here and below, we assume that $\mu < 0$ ($x<0.5$).
\bibitem{factor} For the case of G-AFM, this should be multiplied by $1/\sqrt{2}$.
\bibitem{misprint} Misprints in the published version of these equations are
corrected here.
\end{references}
\end{document}